\documentclass[twocolumn]{aastex631}

\newcommand{\Msun}{\, {\rm M}_\odot}
\newcommand{\logMstar}{\log_{10}{(M_\star/{\rm M}_\odot)}}
\newcommand{\logMBH}{\log_{10}{(M_{BH}/{\rm M}_\odot)}}
\newcommand{\logsSFR}{\log_{10}{(sSFR/{\rm yr}^{-1})}}
\newcommand{\distkpc}{\, {\rm ckpc} \, h^{-1}}
\newcommand{\distMpc}{\, {\rm cMpc} \, h^{-1}}
\def\astrid{\texttt{ASTRID} }
\def\astridN{\texttt{ASTRID}}
\def\tng{\texttt{IllustrisTNG} }
\def\tngN{\texttt{IllustrisTNG}}

\begin{document}

\title{Discrepancies Between JWST Observations and Simulations of Quenched Massive Galaxies at $z > 3$: A Comparative Study With IllustrisTNG and ASTRID}

\author[0009-0000-6612-0599]{Emma Jane Weller}
\affiliation{Center for Astrophysics $\vert$ Harvard \& Smithsonian, Cambridge, MA 02138, USA} 
\affiliation{Department of Astronomy, Yale University, New Haven, CT 06511, USA} 
\email{emma.weller@yale.edu}

\author[0000-0001-9879-7780]{Fabio Pacucci}
\affiliation{Center for Astrophysics $\vert$ Harvard \& Smithsonian, Cambridge, MA 02138, USA} 
\affiliation{Black Hole Initiative, Harvard University, Cambridge, MA 02138, USA}
\email{fabio.pacucci@cfa.harvard.edu}

\author[0000-0001-7899-7195]{Yueying Ni}
\affiliation{Center for Astrophysics $\vert$ Harvard \& Smithsonian, Cambridge, MA 02138, USA}
\email{yueying.ni@cfa.harvard.edu}

\author[0000-0001-6950-1629]{Lars Hernquist}
\affiliation{Center for Astrophysics $\vert$ Harvard \& Smithsonian, Cambridge, MA 02138, USA}

\author[0000-0002-8435-9402]{Minjung Park}
\affiliation{Center for Astrophysics $\vert$ Harvard \& Smithsonian, Cambridge, MA 02138, USA}

\begin{abstract}
Recent JWST observations have uncovered an unexpectedly large population of massive quiescent galaxies at $z>3$. 
Using the cosmological simulations \tng and \astridN, we identify analogous galaxies and investigate their abundance, formation, quenching mechanisms, and post-quenching evolution for stellar masses $9.5 < \logMstar < 12$.
We apply three different quenching definitions and find that both simulations significantly underestimate the comoving number density of quenched massive galaxies at $z \gtrsim 3$ compared to JWST observations by up to $\sim 2$ dex. 
In both simulations, the high-$z$ quenched massive galaxies often host overmassive central black holes above the local $M_{BH}-M_\star$ relation, implying that AGN feedback is key in quenching galaxies in the early Universe. 
The typical quenching timescales for these galaxies are $\sim 200-600$ Myr. 
\tng primarily employs AGN kinetic feedback, while \astrid relies on AGN thermal feedback at $z > 2.3$, which is less effective and has a longer quenching timescale. Although these simulations differ in many aspects, making a direct comparison challenging, our findings suggest the need for improved physical models of AGN feedback in galaxy formation simulations.
At lower stellar masses, the quenched galaxies have denser local environments than the star-forming galaxies, suggesting that environmental quenching helps quench less massive galaxies.
We also study the post-quenching evolution of the high-$z$ massive quiescent galaxies and find that many experience subsequent reactivation of star formation, evolving into primary progenitors of $z=0$ brightest cluster galaxies.
\end{abstract}

\keywords{Quenched galaxies (2016)  --- Early universe (435) --- Hydrodynamical simulations (767) --- Galaxy evolution (594) --- Star formation (1569)}


\section{Introduction} \label{sec:intro}

The James Webb Space Telescope (JWST) has revealed several unexpected properties of high-redshift galaxies and their population of supermassive black holes.

First, overmassive black holes have been detected at $z > 4$ (see, e.g., \citealt{Harikane_2023, Maiolino_2023_new, Yue_2023, Kocevski_2024, Natarajan_2024, Juodzbalis_2024, Taylor_2024_CEERS}), with masses $10-100$ times higher than what is predicted from the local $M_{BH}-M_\star$ relation. These observations challenged standard scaling relations at high redshift and implied the important role played by AGN feedback in quenching galaxies at $z > 4$, especially because their hosts are compact \citep{Pacucci_2023_JWST, Pacucci_2024_LRD}.

Second, and possibly connected \citep{Pacucci_2022_degeneracy, Pacucci_Loeb_2024}, JWST observations have revealed a surprisingly large population of quenched (i.e., not star-forming) massive galaxies at $z>3$ (see, e.g., \citealt{Carnall_2023, Long_2023, Valentino_2023}). Some of these quiescent galaxies, at $3<z<5$, have an estimated stellar mass of $\logMstar > 11$ \citep{Carnall_2024_ultramassive}.
Remarkably, these galaxies can form half their stars as early as $z \sim 11$ \citep{Glazebrook_2023, Park_2024}.

Various classes of quenching mechanisms can be identified. First, AGN and stellar feedback (see, e.g., \citealt{Silk_1998, DiMatteo_2005, Springel_2005, Hopkins_2006, Schaye_2015, Feldmann_2015}) are mechanisms in which the energy output of the central black hole and supernova explosions can prevent the formation of new stars. 
Second, ``dry mergers" between galaxies containing little or no cold gas (see, e.g., \citealt{Barnes_1996, Hopkins_2006, Zolotov_2015, Belli_2019}) can also lead to the formation of massive quiescent galaxies. 
Lastly, environmental quenching is characteristic of galaxies embedded in large clusters where the galactic halo is too hot to be penetrated by cold streams of gas (see, e.g., \citealt{Gunn_1972, Dressler_1980, Mihos_1996, Kawata_2008, vdB_2008, Peng_2010, Donnari_2021_mechanisms}).

Quenching of massive galaxies can be roughly classified into two categories: ``rapid" quenching, occurring on the order of a few hundred million years, and ``slow'' quenching, taking place on timescales longer than 1 billion years (see, e.g., \citealt{Wu_2018, Belli_2019}).
For $z>3$, when the Universe is only $\sim 2$ Gyr old, quiescent galaxies must have undergone quenching in relatively short timescales. 
However, models and simulations struggle to match the observed number density of high-$z$ quenched massive galaxies (see, e.g., \citealt{Schreiber_2018, Cecchi_2019, Girelli_2019, deGraaff_2024}), indicating that the employed physical mechanisms for quenching are insufficient. Recent simulation suites (see, e.g., \citealt{Donnari_2021, Kimmig_2023, RSR_2023}) are becoming increasingly successful in reproducing quenched fractions similar to those observed.

AGN feedback is considered one of the most important mechanisms for the rapid quenching of high-$z$ massive quiescent galaxies (see, e.g., \citealt{Weinberger_2018, Hartley_2023}).
In this study, we use two different simulations, \tng \citep{Weinberger_2017_methods, Pillepich_2018_methods} and \astrid \citep{Bird2022, Ni2022}, with different mechanisms of AGN feedback, to understand the formation, quenching process, and post-quenching evolution of quenched massive galaxies at $z>3$. 

Our paper is organized as follows. In Section \ref{sec:simulations}, we describe the simulations used; in Sec.~\ref{sec:abundance}, we describe the abundance of quenched galaxies and compare them with JWST observations. Sec.~\ref{sec:evolution} describes the redshift evolution of the quenched massive galaxy population investigated in the simulations. 
Sec.~\ref{sec:discussion} provides a broader discussion of our results and concludes our paper.

\section{Simulations} \label{sec:simulations}
This Section briefly summarizes the cosmological simulations used: \tng (the TNG100 and TNG300 volumes) and \astridN. More details can be found in the original papers referenced below.

\tng \citep{Weinberger_2017_methods, Marinacci_2018, Naiman_2018, Nelson_2018, Pillepich_2018, Pillepich_2018_methods, Springel_2018} includes physical models for galaxy formation based on sub-grid models for star formation, stellar evolution and galactic winds, as well as supermassive black hole (SMBH) seeding, merging, accretion, and feedback. The two simulations used, TNG100 and TNG300, have box sizes of $75 \distMpc$ and $205 \distMpc$ and baryon mass resolutions of $1.4 \times 10^6 \Msun$ and $1.1 \times 10^7 \Msun$, respectively. Feedback from galactic winds is kinetic and implemented via hydrodynamically decoupled wind particles. Energy and mass loading factors are prescribed based on local velocity dispersion and metallicity.

For a given SMBH, AGN feedback operates in either a high-accretion quasar mode or a low-accretion radio mode (sometimes called jet mode), depending on the SMBH mass and the Eddington ratio ($\dot{M}_{BH} / \dot{M}_{\rm Edd}$). Below is a summary of the AGN feedback model. For more details, see \cite{Weinberger_2017_methods} and \cite{Nelson_2018}.

An SMBH enters the low-accretion mode when:
\begin{equation} \label{eqn:feedback_threshold}
    \frac{\dot{M}_{BH}}{\dot{M}_{\rm Edd}} < \chi (M_{BH}),
\end{equation}
where $\chi (M_{BH})$ is a threshold that increases with the black hole mass and has a maximum value of $0.1$.

The feedback energy in the high-accretion mode is given by:
\begin{equation} \label{eqn:thermal_feedback}
    \Delta \dot{E}_{\rm high} = \epsilon_{\rm f, th} \epsilon_{\rm r} \dot{M}_{BH} c^2,
\end{equation}
where $\epsilon_{\rm f, th} = 0.1$ is the coupling efficiency and $\epsilon_{\rm r} = 0.2$ is the radiative efficiency. The energy is injected into the nearby gas cells as pure thermal energy, so we refer to this mode as AGN thermal feedback.

The feedback energy in the low-accretion mode is given by:
\begin{equation} \label{eqn:kinetic_feedback}
    \Delta \dot{E}_{\rm low} = \epsilon_{\rm f, kin} \dot{M}_{BH} c^2,
\end{equation}
where the coupling efficiency $\epsilon_{\rm f, kin}$ has a maximum value of $0.2$ but is lower when the gas density around the black hole is low. The energy is injected into the nearby gas cells as pure kinetic energy, so we refer to this mode as AGN kinetic feedback. It is more efficient than thermal feedback at ejecting gas and quenching star formation in massive galaxies \citep{Weinberger_2017_methods, Nelson_2018}.

\astrid \citep{Bird2022, Ni2022} is a large-volume cosmological hydrodynamic simulation of box size $250 \distMpc$ and an initial particle load of $5500^3$ dark matter particles and the same number of gas particles, currently run to $z=0.5$. It has a baryon mass resolution of $1.9 \times 10^6 \Msun$. In the \astrid simulation, the gas cools via primordial radiative cooling \citep{Katz1996} and metal line cooling. The evolution of gas and stellar metallicities follows the prescriptions by \cite{Vogelsberger2014}. The simulation adopts patchy re-ionization \citep{Battaglia2013}, with the ionizing UV background from \cite{FG2020} employed along with gas self-shielding \citep{Rahmati2013}. Star formation is based on a multi-phase model as described in \cite{SH03}, accounting for the influence of molecular hydrogen \citep{Krumholz2011}. Type II supernova wind feedback is incorporated into the simulation following \cite{Okamoto2010}, with wind speeds proportional to the local one-dimensional dark matter velocity dispersion.

\astrid models black hole seeding, accretion, feedback, and dynamics as detailed in \cite{Ni2022} and \cite{Ni_2024}. At $z>2.3$, the AGN feedback operates purely in the thermal form. At $z < 2.3$, AGN feedback can operate in either the thermal or kinetic form, according to the threshold on the Eddington ratio $\chi(M_{BH})$. This is around the epoch where AGN kinetic feedback begins to play a significant role in galaxy evolution. The kinetic and thermal feedback modes follow the same prescription as in \tng (see eqns. \ref{eqn:feedback_threshold} - \ref{eqn:kinetic_feedback}), but with different coefficients: $\epsilon_r = 0.1$, $\epsilon_{\rm f, th} = 0.05$, $\epsilon_{\rm f, kin}$ has a maximum value of $0.05$, and $\chi(M_{BH})$ has a maximum value of $0.05$. The differences in AGN feedback prescriptions between \tng and \astridN, and in particular the lack of AGN kinetic feedback in \astrid at high redshift, allow us to explore the role of AGN feedback in high-redshift quiescent galaxies.

\section{Abundance of quenched massive galaxies} \label{sec:abundance}

In this Section, we present our definitions of quenched massive galaxies, compare the comoving number densities of these galaxies in simulations to recent observational results, and study the stellar mass distribution of quenched versus star-forming massive galaxies.

\subsection{Definitions} \label{sec:definitions}

\begin{figure}
    \includegraphics[width=\columnwidth]{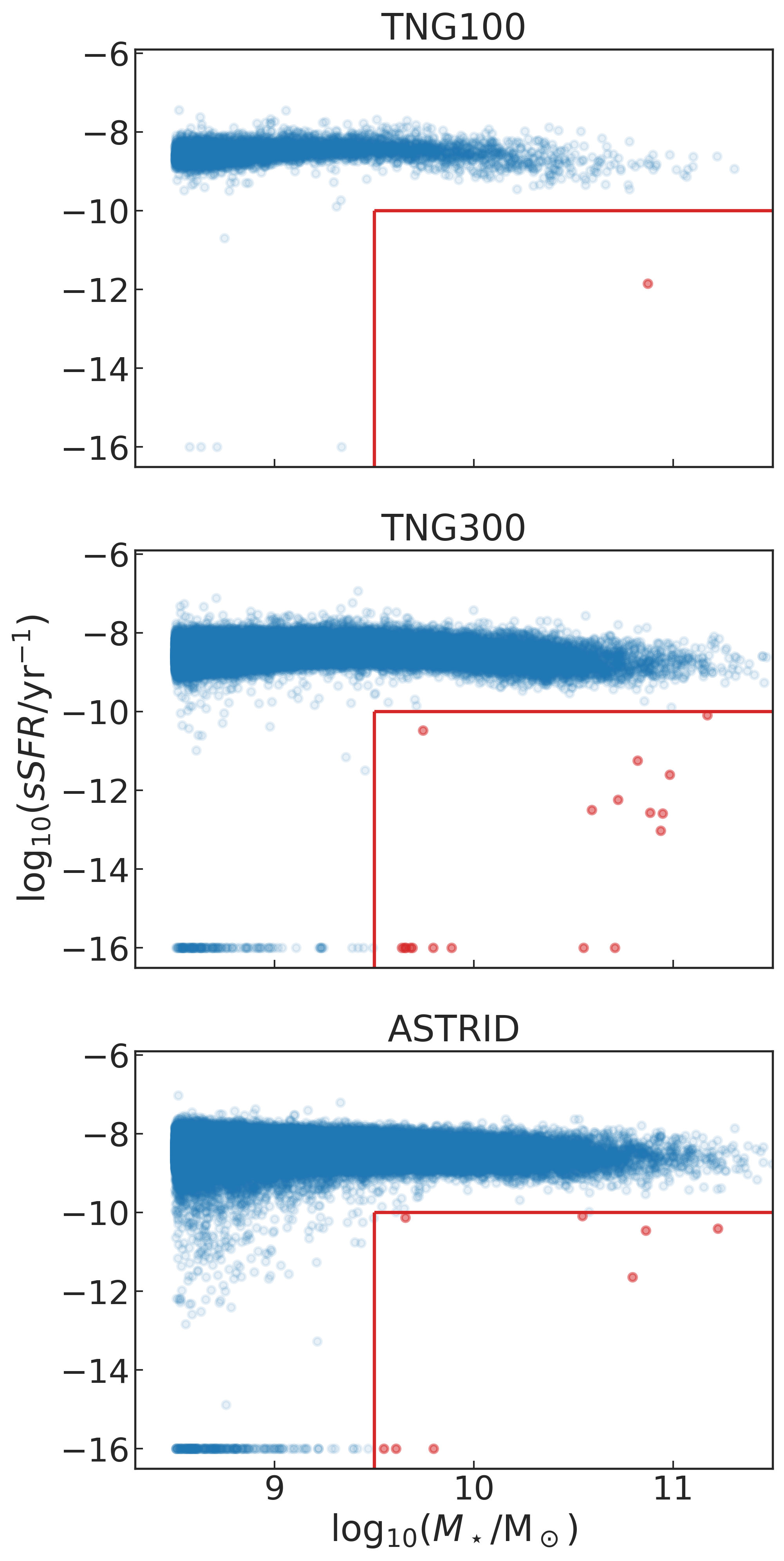}
    \caption{sSFR vs. stellar mass for galaxies at $z=4$ in TNG100 and TNG300 with $\logMstar > 8.5$. The red box marks the region of galaxies that are massive ($\logMstar > 9.5$) and quenched ($\logsSFR < -10$, according to our sSFR limit definition of quenched galaxies). Galaxies within this region, colored in red, are offset from the rest of the population. For visualization purposes, galaxies with $sSFR = 0$ are artificially set to $\logsSFR = -16$.}
    \label{fig:spsfr_defn}
\end{figure}

\begin{figure}
    \includegraphics[width=\columnwidth]{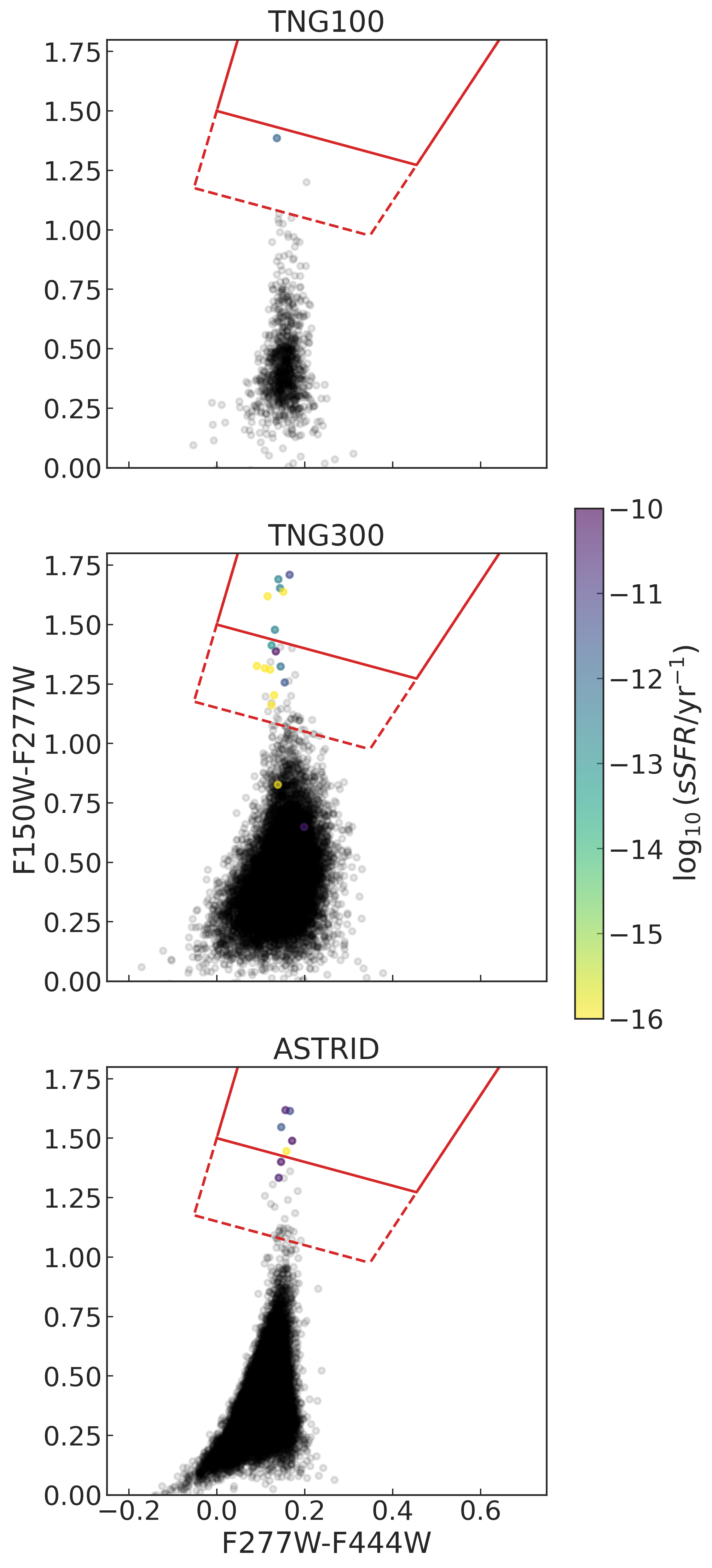}
    \caption{$F150W - F277W$ vs. $F277W - F444W$ colors for massive galaxies at $z=4$ in TNG100 and TNG300. The red lines enclose the regions that meet the short and long wedge criteria for quenched galaxies from \cite{Long_2023}. The dashed lines mark the extension of the long wedge past the short wedge. Galaxies are colored according to their sSFR, and all galaxies with $\logsSFR > -10$ are black. The population of galaxies in the long wedge approximately matches the population with $\logsSFR < -10$. As in Fig.~\ref{fig:spsfr_defn}, galaxies with $sSFR = 0$ are set to $\logsSFR = -16$} for visualization purposes.
    \label{fig:colormag_defn}
\end{figure}

We focus on massive galaxies with stellar masses $\logMstar > 9.5$. Throughout this work, the stellar mass and star formation rate (SFR) are calculated within twice the stellar half-mass radius of the galaxy. As noted in Sec.~\ref{sec:simulations}, the baryon mass resolution is $\sim 10^6 \Msun$ in TNG100 and \astrid and $\sim 10^7 \Msun$ in TNG300. This means that for $\logMstar \lesssim 9$, galaxies are only resolved with $\sim 100 - 1000$ stellar particles. This can lead to underestimated SFR \citep{Pillepich_2018}, so many of these smaller galaxies are artificially quenched, as seen in the low-mass end of Fig. \ref{fig:spsfr_defn}, which shows sSFR vs. stellar mass for galaxies at $z=4$ in the three simulations. We apply three different definitions of quenched galaxies:

\begin{itemize}
    \item \textbf{sSFR limit:} A galaxy is quenched if its specific star formation rate (sSFR, star formation rate per unit stellar mass) is $\logsSFR < -10$ \citep{Donnari_2021, Valentino_2023}. See Fig.~\ref{fig:spsfr_defn} for an example of this selection criterion for massive galaxies at $z=4$ in the three simulations (see, e.g., \citealt{Hartley_2023}). 
    \item \textbf{Deviation from main sequence:} We divided the stellar mass range into bins of $0.2 \, \rm dex$ and calculated the median SFR for each bin. The bin medians define the ``main sequence'' of star formation. A galaxy is quenched if its SFR lies more than 1 dex below the median for its stellar mass bin (see, e.g., \citealt{Donnari_2021, ENelson_2021}). This definition agrees well with the sSFR definition, especially in TNG100 and TNG300 (see Fig. \ref{fig:number_density}).
    \item \textbf{Color selection:} We used the \texttt{Flexible Stellar Population Synthesis} (FSPS) Python library \citep{Conroy_2009, Conroy_2010} to calculate the magnitude of the galaxies in the JWST Near Infrared Camera (NIRCam) photometric bands $F150W,$ $F227W,$ and $F444W.$ For each stellar particle (which represents many stars) in a galaxy, we used FSPS to create a stellar population with the particle's age and metallicity and the galaxy's redshift. We assumed a Chabrier initial mass function with the total mass normalized to the mass of the particle. FSPS then produced a spectral energy distribution for the stellar population and used this to calculate the apparent magnitude in the specified bands. Next, we computed the combined magnitude in each band for all the particles within twice the galaxy's stellar half-mass radius. We note that we did not account for dust attenuation. After finding the magnitudes in the three JWST bands, we calculated the $F150W-F277W$ and $F277W-F444W$ colors and identified quenched galaxies using the ``long wedge'' selection criteria from \cite{Long_2023} (developed for $3 \lesssim z \lesssim 6$)
    \begin{equation}
        (F150W - F227W) < 1.5 + 6.25 \times (F227W - F444W)
    \end{equation}
    \begin{equation}
        (F150W - F227W) > 1.15 - 0.5 \times (F227W - F444W)
    \end{equation}
    \begin{equation}
        (F150W - F227W) > 2.8 \times (F227W - F444W)
    \end{equation}
    Figure~\ref{fig:colormag_defn} shows an example of the short and long wedge selection criteria for massive galaxies  at $z=4$ in the three simulations. The color selection definition of quenched galaxies agrees approximately with the sSFR limit definition (see also Fig. \ref{fig:number_density}). We primarily use the sSFR limit definition in this paper (starting in Sec.~\ref{sec:stellar_mass_range}), as sSFR is a property that is well-defined in the simulations.

\end{itemize}

\subsection{Number density} 
\label{sec:number_density}

\begin{figure*}
    \includegraphics[width=2\columnwidth]{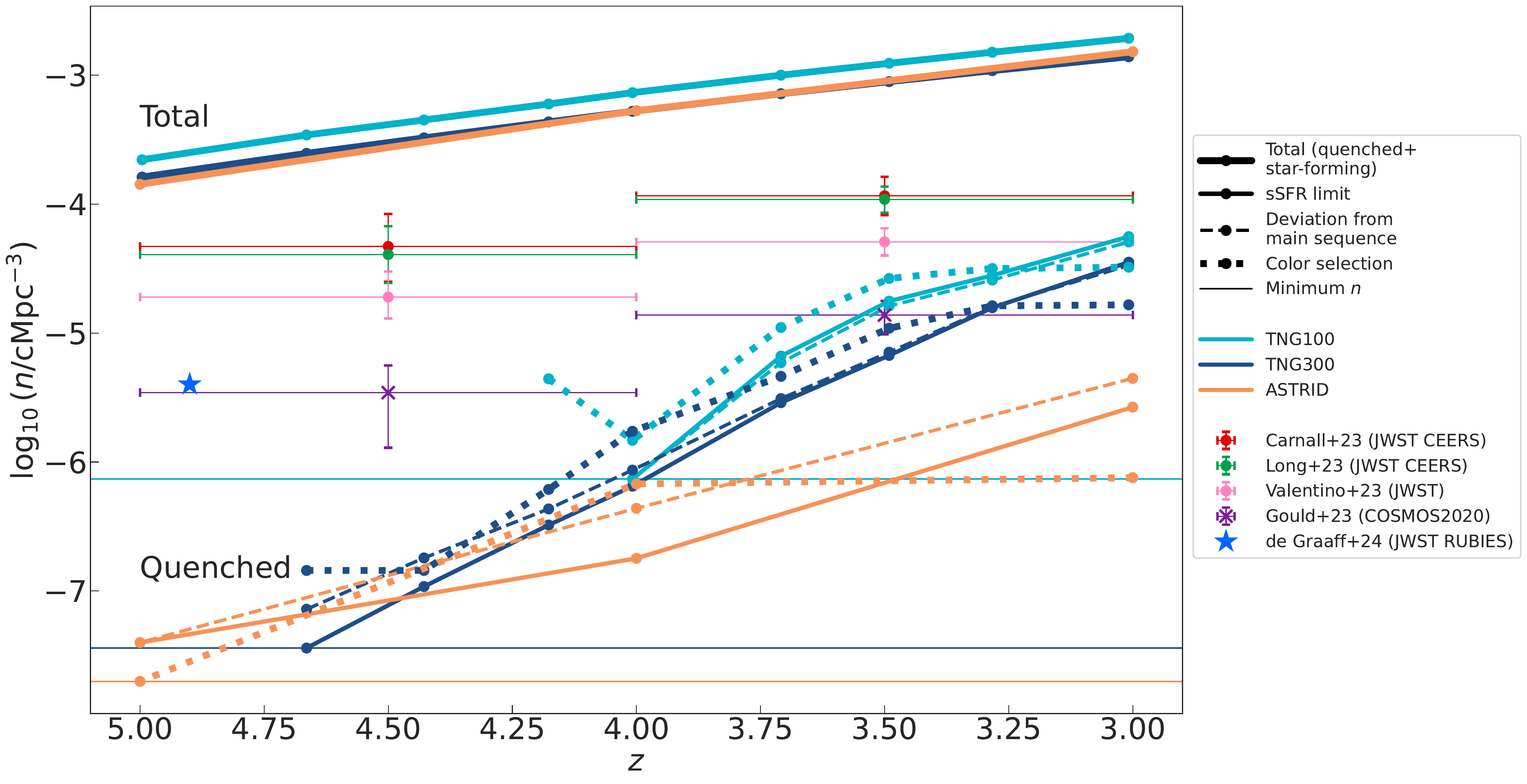}
    \caption{Comoving number density of quenched massive galaxies in TNG100, TNG300, and \texttt{ASTRID}, as a function of redshift, compared to recent observational estimates \citep{Carnall_2023, Long_2023, Valentino_2023, Gould_2023, deGraaff_2024}. In each simulation, we use three different definitions of quenched galaxies. We also show the total number density of massive galaxies for comparison. The three definitions agree reasonably well, but the simulations produce significantly lower quenched number densities than observations. JWST has the highest values, and \astrid has the lowest. The number density decreases with increasing redshift. Note that the lines stop before $z=5$ if their number densities fall to zero (for example, there are no quenched massive galaxies in TNG100 at $z>4$).}
    \label{fig:number_density}
\end{figure*}

We calculate the comoving number density of galaxies at each snapshot between $z=3$ and $z=5$ in TNG100, TNG300, and \texttt{ASTRID}. 
We limit our study to $z<5$ due to the limited volumes of the simulations, as they contain essentially no earlier quenched massive galaxies. 
We begin by counting the number of quenched massive galaxies using each of the three definitions described in Sec.~\ref{sec:definitions}. 
We also count the total number of massive galaxies. We then convert to comoving number densities using the volumes of the simulation boxes: $(75 \distMpc)^3$, $(205 \distMpc)^3$, and $(250 \distMpc)^3$ for TNG100, TNG300, and \texttt{ASTRID}, respectively. 
The results are shown in Fig.~\ref{fig:number_density}, compared to observational comoving number densities from four recent studies using JWST and COSMOS2020 \citep{Carnall_2023, Long_2023, Valentino_2023, Gould_2023}. The following paragraph summarizes the methods that the four observational studies used to select their samples of candidate quenched massive galaxies at $3 < z < 5$.

\cite{Carnall_2023} used Near Infrared Camera (NIRCam) photometric data from the JWST Cosmic Evolution Early Release Science survey (CEERS, \citealt{Finkelstein_2023_CEERS}). This study found 15 quenched massive galaxies with $9.63 \leq \logMstar \leq 11.53$. They consider a galaxy to be quenched if $sSFR < 0.2/t_{obs}$, where $t_{obs}$ is the age of the Universe at the redshift of the galaxy (note that $0.2/t_{obs} \approx 10^{-10} \, {\rm yr}^{-1}$ at $z=3$, so this is similar to our sSFR limit definition over the studied redshifts). \cite{Long_2023} selected 36 quenched galaxies with $8.8 \leq \logMstar \leq 11.3$ by applying their new empirical color selection method to NIRCam data from CEERS. \cite{Valentino_2023} applied color selection to identify $\sim 80$ quenched massive galaxies with $9.5 \leq \logMstar \lesssim 12$. They used 11 fields imaged by NIRCam, the Slitless Spectrograph (NIRISS), and the Mid-Infrared Instrument (MIRI). \cite{Gould_2023} used color selection to identify a sample of 230 quenched massive galaxies with $10.6 \leq \logMstar \lesssim 12$ in the Cosmic Evolution Survey 2020 catalog (COSMOS2020).

Figure~\ref{fig:number_density} shows that the simulations significantly underestimate the number density of quenched massive galaxies compared to the observational results, especially the JWST results. The difference is more extreme for \astrid and at higher redshifts, and at $z = 4.5$ it reaches $\sim 2 \, \rm dex$. This $2 \, {\rm dex}$ discrepancy was also found by \cite{deGraaff_2024}, who reported the spectroscopic confirmation of a massive quiescent galaxy at $z=4.9$ using the JWST Red Unknowns: Bright Infrared Extragalactic Survey (RUBIES, \citealt{deGraaff_2024_RUBIES}). At the time, this was the highest-redshift spectroscopically confirmed massive quiescent galaxy. The estimated quenched number density based on this galaxy is included in Fig.~\ref{fig:number_density}. 

As noted in Sec.~\ref{sec:definitions}, we also see from Fig. \ref{fig:number_density} that our three definitions of quenched galaxies agree reasonably well, especially in \tngN.

\subsection{Stellar mass distribution} \label{sec:stellar_mass_range}

\begin{figure}
    \includegraphics[width=\columnwidth]{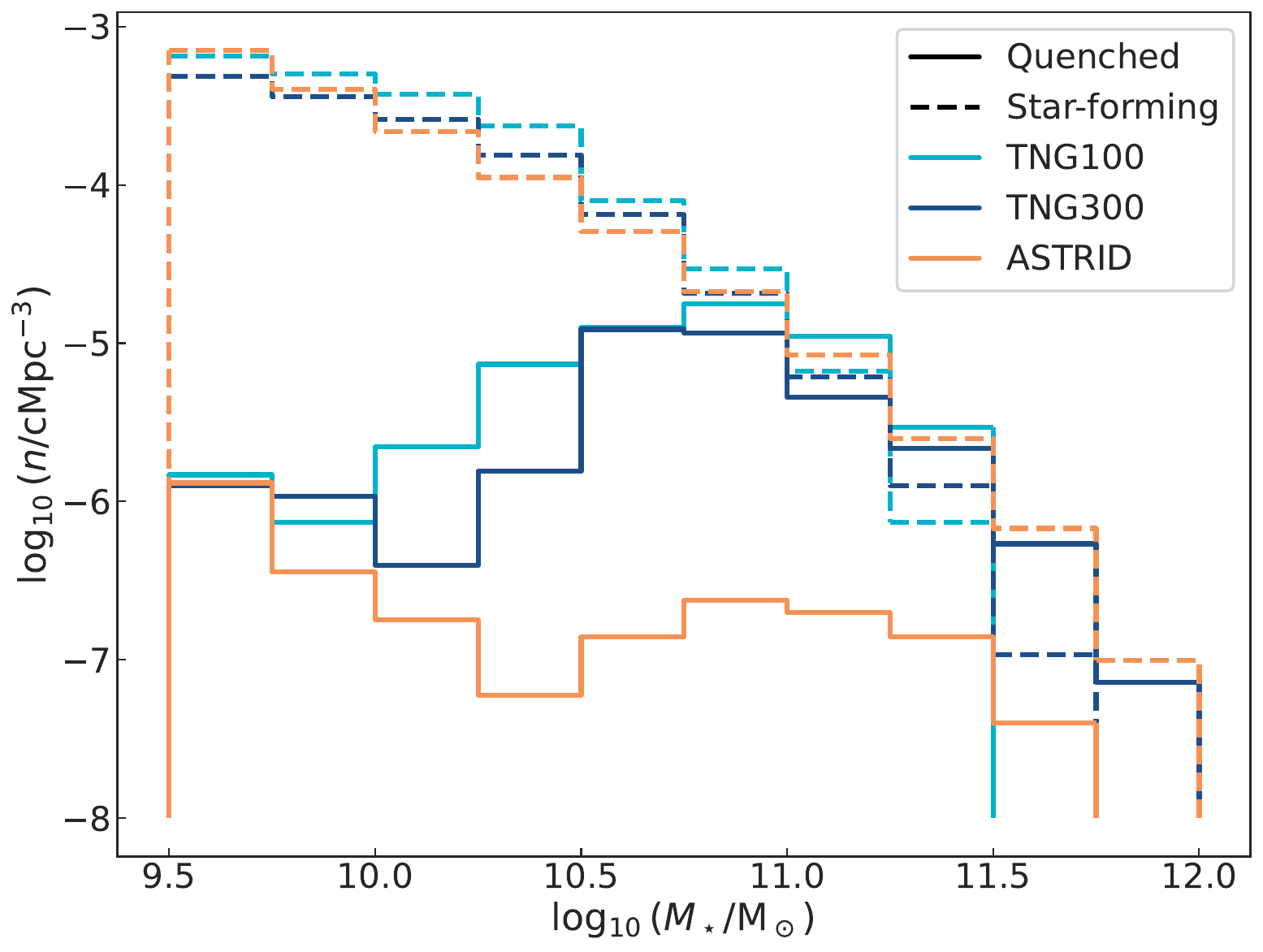}
    \caption{Comoving number density of quenched and star-forming massive galaxies at $z=3$ in all three simulations. Values are calculated in stellar mass bins of width $0.25 \, {\rm dex}$. The star-forming number density decreases with the stellar mass, while the quenched number density has a peak (smaller in \texttt{ASTRID}) at $\logMstar \sim 11$.}
    \label{fig:stmass_hist}
\end{figure}

Figure~\ref{fig:spsfr_defn} showed that quenched galaxies tend to occupy the parameter space with $\logMstar \gg 9.5$.
Figure~\ref{fig:stmass_hist} now displays the distribution of stellar masses for quenched and star-forming massive galaxies in all three simulations at $z=3$. 
Note that here and throughout the remainder of the paper, ``quenched'' refers to galaxies that meet the sSFR limit definition.

The star-forming comoving number densities are similar in all three simulations and decrease with stellar mass. The quenched comoving number density, on the other hand, peaks at $\logMstar \sim 11$ in TNG100 and TNG300. 
In \astridN, there appear to be two populations of quenched galaxies, split around $\logMstar \sim 10.25$. The lower-mass population follows a similar distribution to the star-forming galaxies, while the higher-mass population follows a similar distribution to the quenched galaxies in \tngN. We will discuss these two populations in Sec.~\ref{sec:mechanisms}.

When comparing quenched and star-forming galaxies, we must consider that the quenched galaxies have higher average stellar mass. We use binning to separate effects that only depend on stellar mass.

\subsection{$M_{BH}$-$M_\star$ relation} \label{sec:mbh-mstar}

\begin{figure*}
\centering
    \includegraphics[width=2.1\columnwidth]{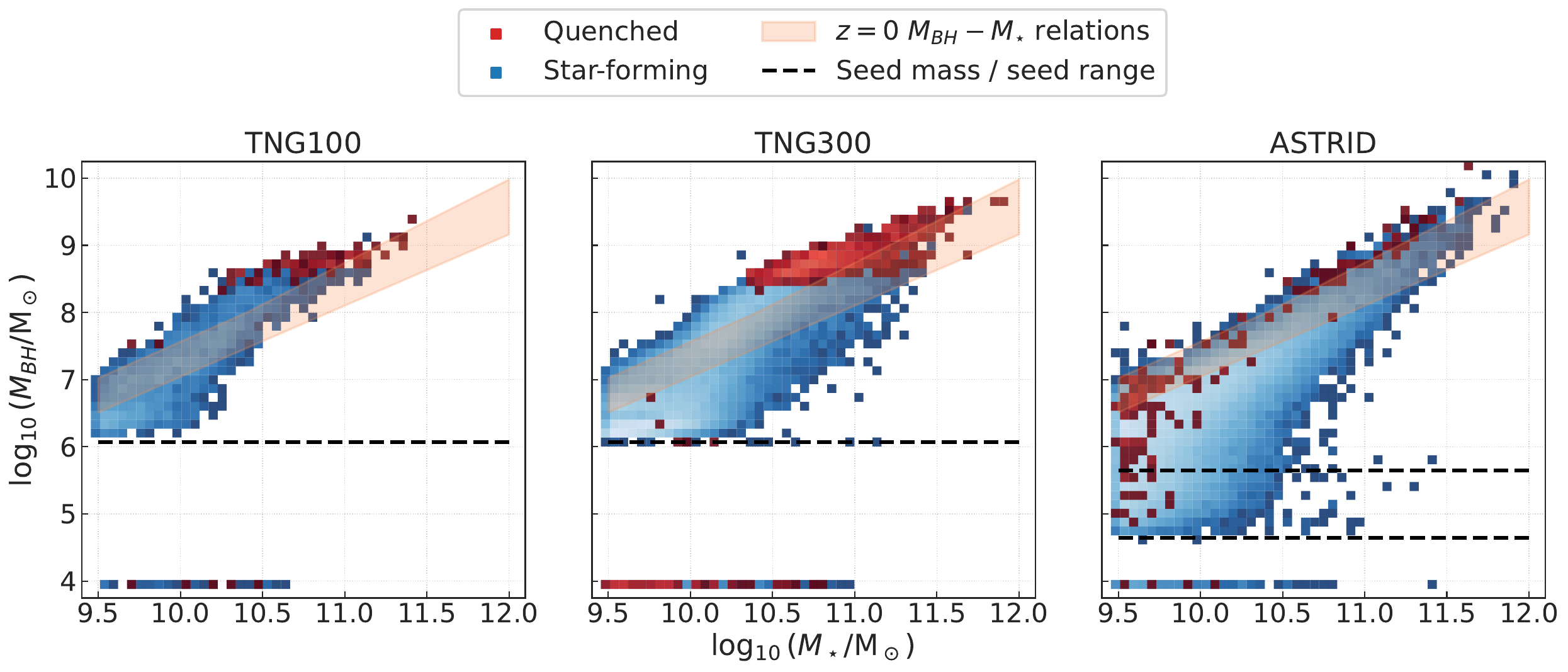}
    \caption{2D histogram of central black hole mass vs. stellar mass for quenched and star-forming galaxies at $z=3$ in all three simulations. Quenched galaxies tend to lie on the upper edge of the distribution at the high stellar masses, but can have lighter SMBHs (or no SMBHs) at lower stellar masses. We mark the range of empirical $z=0$ $M_{BH} - M_\star$ relations from \cite{Haring_Rix_2004}, \cite{McConnell_Ma_2013}, and \cite{Kormendy_Ho_2013}, and the seed masses of each simulation. For visualization purposes, galaxies with $M_{BH} = 0$ are positioned at $\logMBH = 4$. Note that the quenched galaxy distribution is overlaid on top of the star-forming galaxy distribution.}
    \label{fig:Mbh-Mstar}
\end{figure*}

Figure~\ref{fig:Mbh-Mstar} displays the $M_{BH} - M_\star$ relation for quenched and star-forming massive galaxies at $z=3$ in all three simulations. In \tngN, each SMBH is pinned to the center (location of minimum gravitational potential energy) of its host halo, so we take the most massive black hole in the galaxy as the central black hole. In \astridN, because there are wandering massive black holes (see, e.g., \citealt{DiMatteo_2022, Weller_2023}), we use the most massive SMBH within twice the stellar half-mass radius. Our selected central black holes are close to their galactic centers. For all central black holes in the three simulations at $z=3$ with $\logMBH > 8$, we calculate the distance from the location of minimum gravitational potential energy. We find that this distance is $<0.5 \distkpc$ for $97\%$, $83\%$, and $96\%$ of the central black holes in TNG100, TNG300, and \astridN, respectively.

In all three simulations, we see that at the high end of the stellar mass range, quenched galaxies lie on the upper edge of the distribution. At the lower end of the stellar mass range, many galaxies have smaller central SMBHs or none at all. This suggests that AGN feedback may drive quenching at the highest stellar masses, while at lower masses, galaxies are quenched via other mechanisms like environmental effects. This will be discussed in Sec.~\ref{sec:environment}.

Our results are in agreement with \cite{Szpila_2024}, who find that high-redshift quenched massive galaxies in the \texttt{SIMBA-C} simulation have higher $M_{BH}/M_\star$ ratios than their star-forming counterparts (all of the quenched galaxies have $M_{BH}/M_\star > 10^{-3}$). Similarly, \cite{Kurinchi_2023} show that in \tngN, SMBHs in quenched galaxies are seeded earlier and grow faster than in star-forming galaxies.

We note that the median black hole mass among quenched galaxies is smaller in \astrid than in \tng ($10^{6.9} \Msun$ in \astrid compared to $10^{8.7} \Msun$ in both TNG100 and TNG300). This makes sense, as \tng has a higher seed mass than \astrid and is therefore able to produce larger SMBHs by $z=3$. This mass difference should not have a significant effect on the Bondi accretion rates, as the gas density is low around central SMBHs in quenched galaxies.

\section{Evolution of quenched massive galaxies} 
\label{sec:evolution}
In this Section, we study the evolution of high-redshift quenched massive galaxies compared to their star-forming counterparts. 

\subsection{Formation} 
\label{sec:formation}

We studied the formation of quenched massive galaxies in \tng by tracing them through time via their merger trees. We do not complete this analysis in \astrid because its subhalo catalog does not have sufficient time resolution. We define the formation time of a galaxy as the time at which its stellar mass exceeds $10^8 \, \rm M_\odot$ (i.e., when the galaxy becomes resolved in the simulation). We use this definition because we are primarily interested in tracking when galaxies begin to build up their stellar mass. Note that other works use a different definition of formation time, i.e., the time when half of the final stellar mass is formed or the age of the Universe corresponding to the age of the galaxy \citep{Carnall_2023}. 

We took every quenched massive galaxy at $z = 2$ in TNG100 and TNG300 and traced it back in time along the main progenitor branch to identify the formation time, the time at which it becomes massive (the first snapshot in which $\logMstar > 9.5$), the quenching time (the first snapshot in which $\logsSFR < -10$), and the stellar mass at quenching. We also found the time when AGN kinetic feedback turns on (see Sec.~\ref{sec:mechanisms}). Note that time refers to the age of the Universe (e.g., $13.8 \, \rm Gyr$ at $z=0$).

\begin{figure}
    \includegraphics[width=\columnwidth]{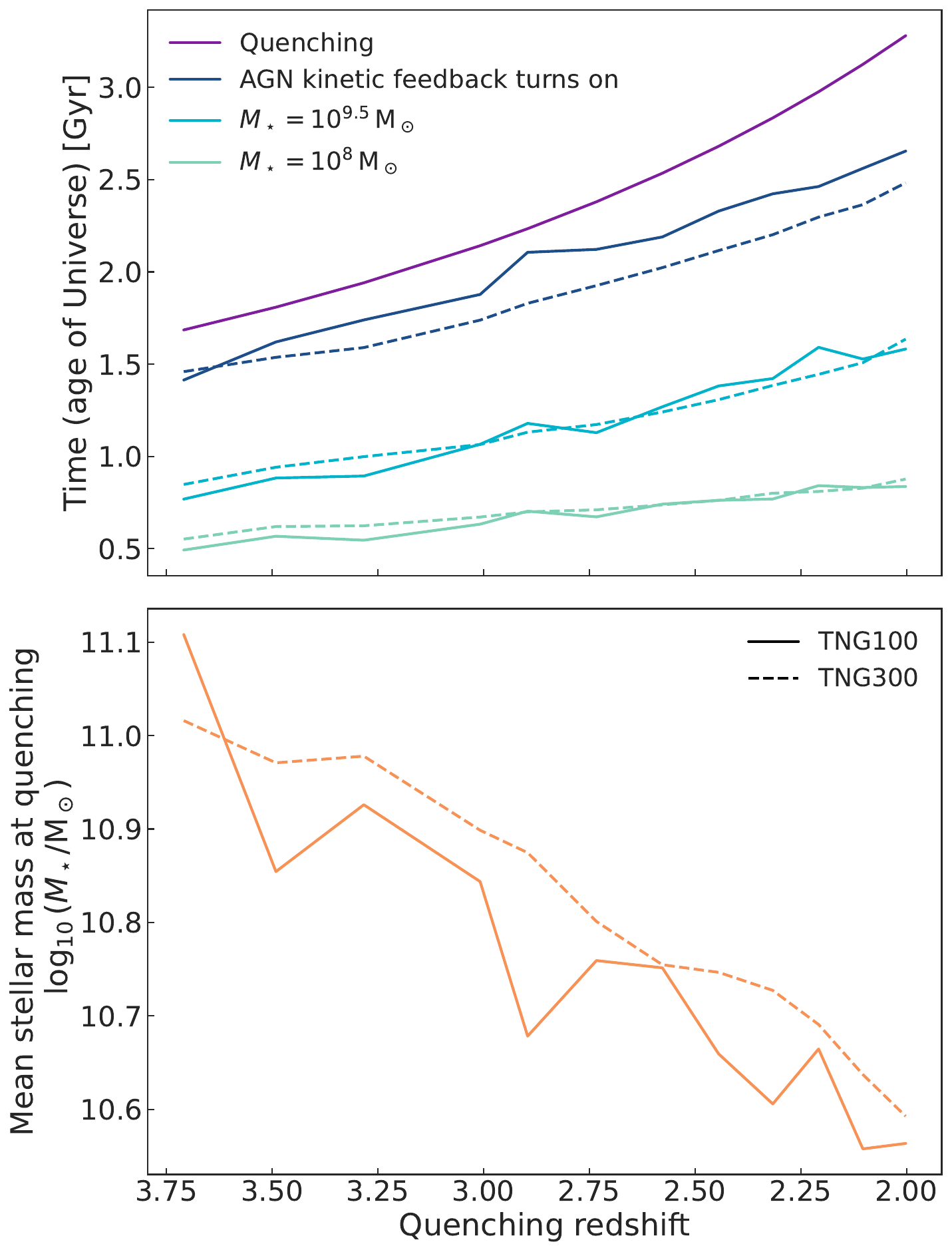}
    \caption{We consider all quenched massive galaxies in TNG100 and TNG300 at $z=2$ and sort them by quenching redshift. \textbf{Upper panel:} We show for each quenching redshift the mean values of the formation time, the time when the galaxies become massive, and the time when AGN kinetic feedback turns on. \textbf{Lower panel:} We show the mean values of the stellar mass at quenching. We see that massive galaxies that quench earlier tend to assemble stars more quickly and that the quenching time is closely correlated with the time at which AGN kinetic feedback turns on.}
    \label{fig:formation}
\end{figure}

We grouped the galaxies by their quenching snapshot and calculated the mean values of the quantities listed above for each snapshot.
We plot these mean values as a function of the quenching redshift for TNG100 and TNG300 in Fig.~\ref{fig:formation}. 
We also show the corresponding quenching time. 
Note that few galaxies have quenching redshifts $z > 3.7$.

On average, the galaxies that quench earlier grow from $\logMstar = 8$ to $\logMstar = 9.5$ more quickly, and they end up more massive at quenching. This fact indicates that massive galaxies that quench earlier tend to assemble stellar mass more quickly.

\subsection{Quenching mechanisms} \label{sec:mechanisms}

In this Section, we investigate how massive galaxies are quenched at high redshifts. 

\subsubsection{AGN kinetic feedback in IllustrisTNG} \label{sec:kinetic_feedback}

\begin{figure}
    \includegraphics[width=\columnwidth]{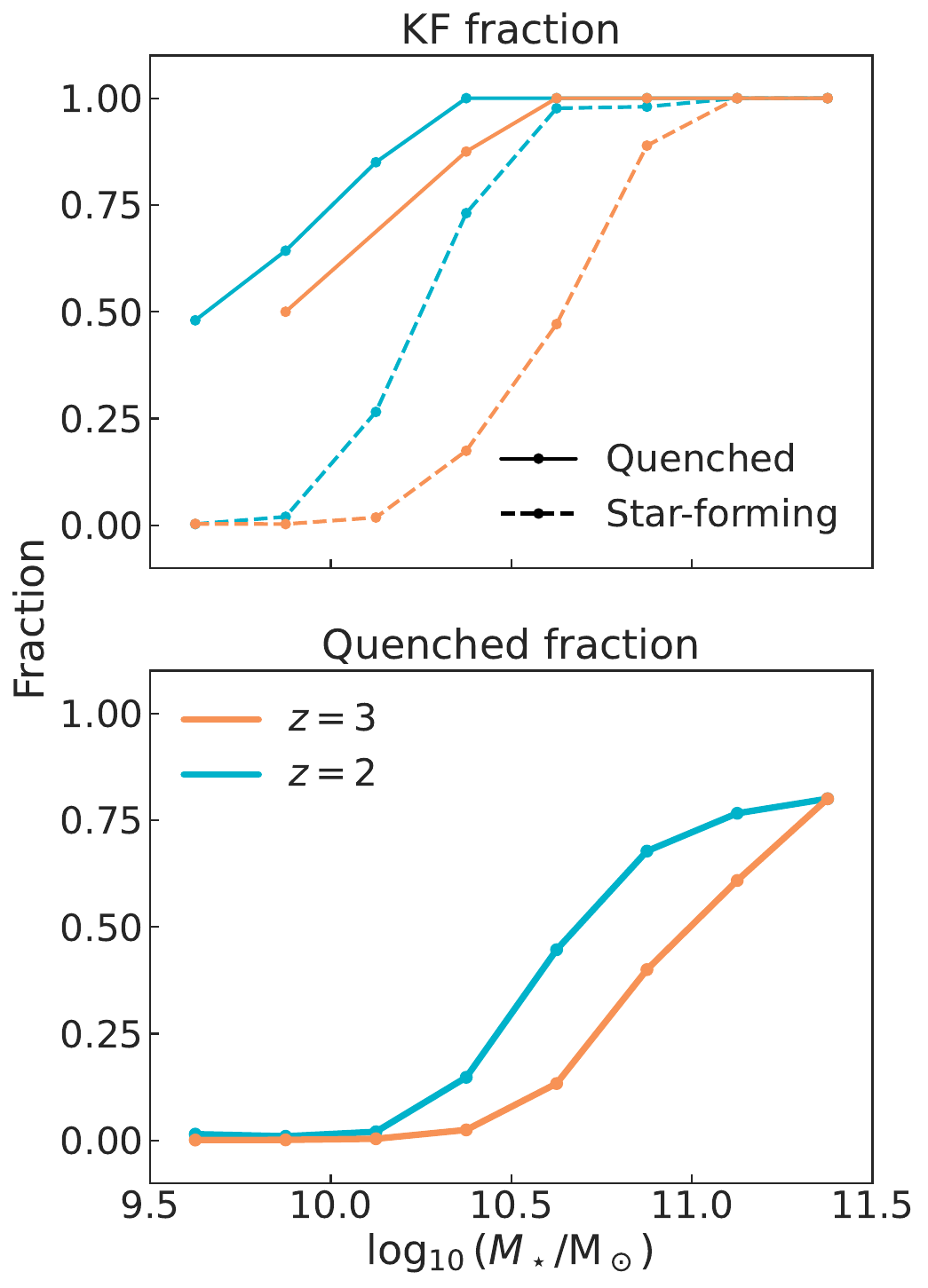}
    \caption{\textbf{Upper panel:} KF fraction (i.e., fraction of galaxies in which AGN kinetic feedback is on) vs. stellar mass for quenched and star-forming massive galaxies in TNG100 at $z=3$ and $z=2$. We divide the stellar mass into bins of width $0.25 \, \rm dex$. Note that the three lowest mass bins are combined for the quenched galaxies at $z=3$. We find that most quenched massive galaxies have AGN kinetic feedback turned on, and nearly all galaxies have kinetic feedback turned on the highest stellar masses. More galaxies have kinetic feedback turned on at $z=2$ than at $z=3$. \textbf{Lower panel:} Fraction of galaxies that are quenched vs. galaxy stellar mass, shown for reference.}
    \label{fig:feedback_frac}
\end{figure}

\begin{figure}
    \includegraphics[width=\columnwidth]{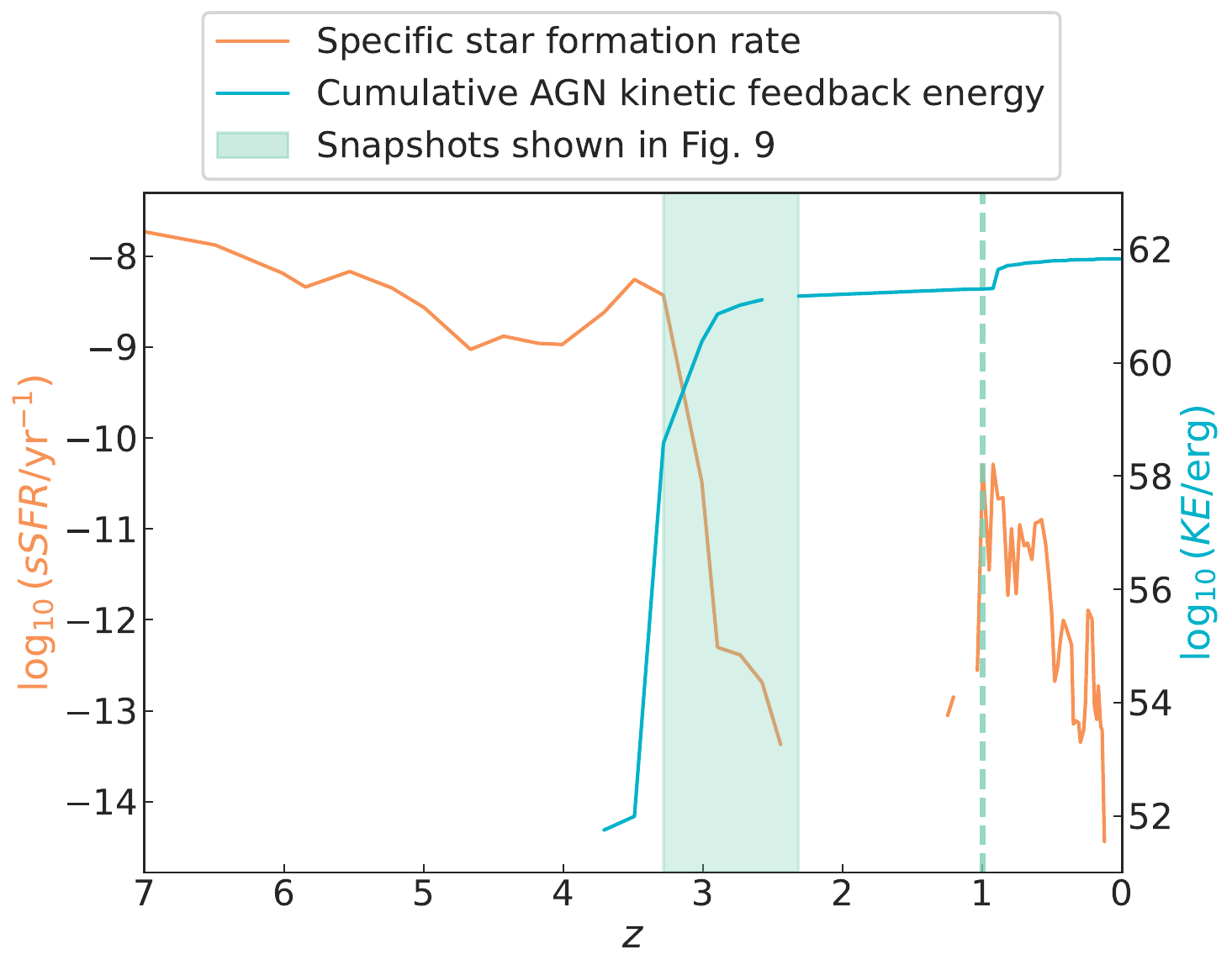}
    \caption{Evolution of an example massive galaxy in TNG100 that quenches at $z = 3$ when AGN kinetic feedback turns on and later reactivates at $z \sim 1$. In Fig.~\ref{fig:TNG100-QG-example}, we illustrate the quenching and reactivation processes by mapping this galaxy at several snapshots near $z=3$ and one snapshot at $z=1$. These snapshots are marked in the plot here by the green shaded region and dashed line, respectively.}
    \label{fig:evolution_plot_fdbk}
\end{figure}

\begin{figure*}
\centering
    \includegraphics[width=2.0\columnwidth]{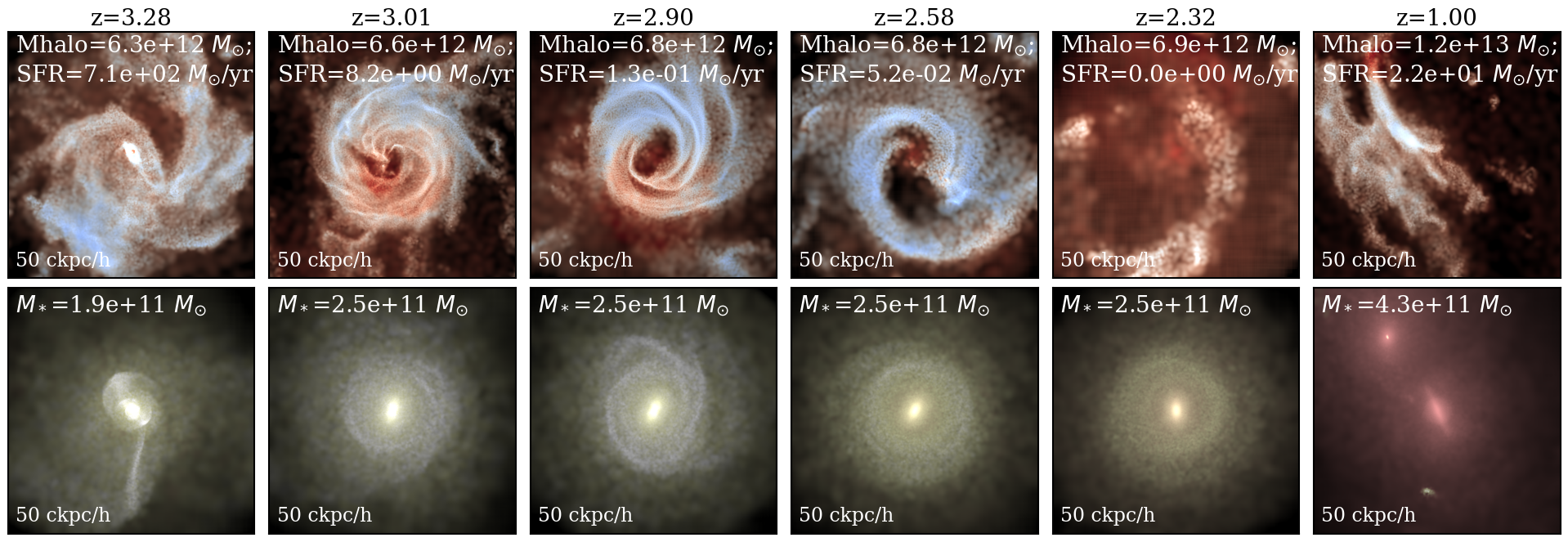}
    \caption{
    Maps of the gas density colored by temperature (top row, redder is hotter) and stellar density colored by age (bottom row, redder is older) of the TNG100 example galaxy from Fig.~\ref{fig:evolution_plot_fdbk}. Brightness indicates the mass density. 
    Each column shows the galaxy at a particular snapshot. For each snapshot, we list the total halo mass and the SFR and stellar mass within twice the stellar half-mass radius. Around the time of quenching (the first five snapshots, running from $z = 3.28 - 2.32$), the gas is heated and expelled from the galaxy's center. At reactivation (which occurs at $z \sim 1$; see the final snapshot), new gas is brought into the galaxy via mergers with surrounding galaxies. All boxes have a side length of $50 \distkpc$.}
    \label{fig:TNG100-QG-example}
\end{figure*}

AGN kinetic feedback plays a pivotal role in quenching massive galaxies in \tng \citep{Weinberger_2017_methods, Weinberger_2018, Nelson_2018, Terrazas_2020, Zinger_2020, Kurinchi_2023}. \cite{Weinberger_2017_methods} show that using kinetic rather than thermal feedback for the low-Eddington radio mode significantly reduces star formation in massive galaxies, leading to better agreement with observations. Returning to Fig.~\ref{fig:formation}, the mean time at which AGN kinetic feedback turns on is closely correlated with the quenching time. This supports the idea that kinetic feedback is responsible for quenching massive galaxies. We also notice that faster stellar mass growth is correlated with an earlier onset of kinetic feedback.

Next, we calculate the so-called ``KF fraction'' for quenched and star-forming massive galaxies in TNG100: the fraction of galaxies in which AGN kinetic feedback has turned on. We assume that kinetic feedback has turned on if the main progenitor of the galaxy has contained a black hole with nonzero cumulative kinetic feedback energy at some point in its history. This allows us to include galaxies quenched by kinetic feedback but which later lost the SMBH (e.g., via interactions with another galaxy). We calculate the KF fraction in stellar mass bins of width $0.25 \, \rm dex$, at $z=3$ and $z=2$.

The upper panel of Fig.~\ref{fig:feedback_frac} displays our results for the KF fraction, which is higher for larger stellar mass, lower redshift, and quenched galaxies. At the high end of the stellar mass range, AGN kinetic feedback has turned on in almost all galaxies. At the low mass end, kinetic feedback has turned on in about half of the quenched galaxies and almost none of the star-forming ones. We also display in the lower panel of Fig.~\ref{fig:feedback_frac} the fraction of galaxies that are quenched as a function of stellar mass.

The quenched galaxies fall primarily at the high end of the studied stellar mass range, and this is where AGN kinetic feedback usually turns on, further supporting the conclusion that it is difficult to quench the most massive galaxies without AGN kinetic feedback. Summing over the entire mass range, we find that $96\%$ of quenched galaxies and just $6\%$ of star-forming galaxies have kinetic feedback turned on at $z=3$. At $z=2$, these values are $96\%$ and $21\%$. However, as noted in Sec.~\ref{sec:mbh-mstar}, lower-mass galaxies can be quenched via other mechanisms such as environmental effects.

Figures~\ref{fig:evolution_plot_fdbk} and \ref{fig:TNG100-QG-example} illustrate the evolution of an example massive galaxy in TNG100 that quenches at $z=3$. At the time of quenching, the mass of its central, most massive SMBH is $\logMBH = 9.3$. In Fig.~\ref{fig:evolution_plot_fdbk}, we display the sSFR and the cumulative AGN kinetic feedback energy of the central SMBH in the galaxy as a function of redshift, for $0 < z < 7$. When AGN kinetic feedback turns on, the sSFR drops rapidly, and the galaxy becomes quenched $0.46 \, {\rm Gyr}$ later at $z=3$. 

In Fig.~\ref{fig:TNG100-QG-example}, we map the gas temperature and stellar age of this example galaxy at snapshots from $z=3.28$ to $z=2.32$ (a period of $0.89 \, {\rm Gyr}$). The gas is heated and ejected from the galaxy's central region -- a signature of AGN kinetic feedback. Quenching occurs at the second snapshot in the figure ($z=3.01$).

\subsubsection{Quenching mechanism in ASTRID} 
\label{sec:astrid_quenching}

\begin{figure}
    \includegraphics[width=\columnwidth]{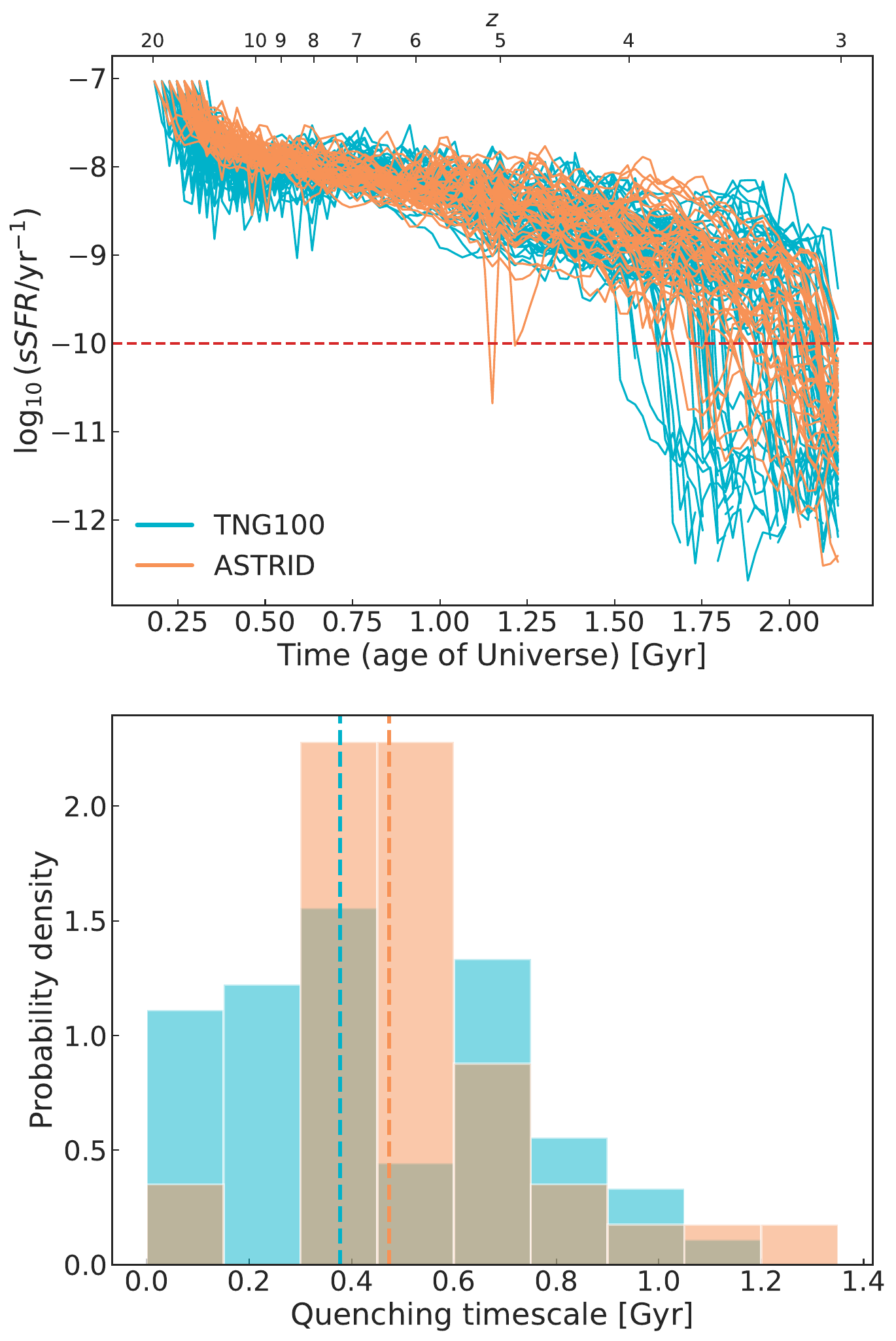}
    \caption{We consider galaxies in TNG100 and \astrid that are quenched and have $\logMstar > 10.5$ at $z=3$. \textbf{Upper panel:} Reconstructed sSFR of the galaxies. They evolve similarly in the two simulations, but several TNG100 galaxies have sSFR drops that are earlier and more abrupt than those seen in \astridN. \textbf{Lower panel:} Histograms of the quenching timescales of the galaxies, normalized to probability density. The median values are marked with dashed lines. The quenching timescales are similar between TNG100 and \astridN, but TNG100 has more galaxies at the shortest timescales.}
    \label{fig:sfh_delay_plot}
\end{figure}

\begin{figure*}
\centering
    \includegraphics[width=2\columnwidth]{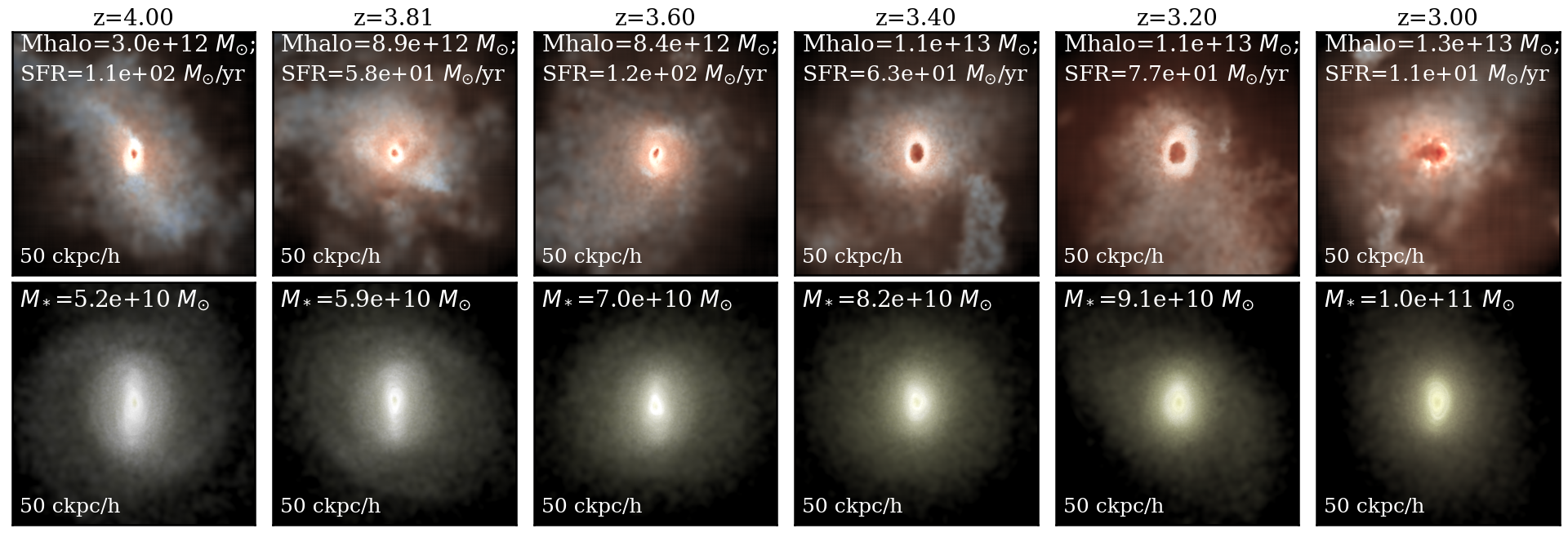}
    \caption{Maps of the gas density colored by temperature (top row, redder is hotter) and stellar density colored by age (bottom row, redder is older) of the \astrid example galaxy, as in Fig.~\ref{fig:TNG100-QG-example}. Brightness corresponds to the mass density. We show the snapshots leading up to quenching at $z=3$. For each snapshot, we list the total halo mass and the reconstructed SFR and stellar mass. Gas is heated at the galaxy's center, but we do not see significant gas expulsion.}
    \label{fig:Astrid-QG-example}
\end{figure*}

At $z>2.3$, \astrid employs AGN feedback only in the thermal mode, so AGN kinetic feedback does not play a role in quenching massive galaxies. Returning to the two populations of quenched \astrid galaxies in Fig.~\ref{fig:stmass_hist}, we first notice that the quenched number density in \astrid approximately matches TNG100 and TNG300 at the low end of the stellar mass range. This is because, as noted in Sec.~\ref{sec:mbh-mstar}, quenching at lower stellar masses is driven not by AGN feedback but by other mechanisms like environmental effects (see Sec.~\ref{sec:environment}). The discrepancy in the quenched number density between \astrid and \tng arises at higher stellar masses. Since quenching at these masses is primarily driven by AGN feedback, the different prescriptions for AGN feedback used in \astrid and \tng may explain this discrepancy. However, we highlight that the simulation suites used also differ in other aspects, so performing an absolute comparison is challenging. 

Due to the low time resolution of the subhalo catalog in \astridN, we investigate the evolution of galaxies over time by reconstructing their star formation histories based on stellar age. We identify the formation time of all stellar particles within twice the stellar half-mass radius at $z=3$, divide the particles into 100 temporal bins, and use their masses to calculate the SFR and sSFR for each bin. In the upper panel of Fig.~\ref{fig:sfh_delay_plot}, we display the reconstructed sSFR for all galaxies in TNG100 and \astrid that are quenched and have $\logMstar > 10.5$ at $z=3$ (this mass cut selects samples of 60 and 38 galaxies, respectively, with comparable stellar masses). While the results are similar between the two simulations, at least up to $\sim 1.5 \, {\rm Gyr}$ ($z \sim 4$), the TNG100 galaxies often exhibit earlier and more abrupt drops in sSFR compared to the \astrid galaxies. These sharp declines are possibly driven by the efficient AGN kinetic feedback in \tngN, as seen in Fig. \ref{fig:evolution_plot_fdbk}, where the sharp drop in sSFR closely corresponds to the onset of AGN kinetic feedback for the TNG100 example galaxy.

The lower panel of Fig.~\ref{fig:sfh_delay_plot} shows histograms of the quenching timescales for these galaxies, calculated as the time between quenching and peak SFR in the reconstructed star formation history. Note that the reconstructed sSFR sometimes goes to zero near the formation time, so we require the quenching time to be after $1.0 \, {\rm Gyr}$. The quenching timescales are similar between the two simulations. However, there are more TNG100 galaxies at the shortest quenching timescales, leading to a slightly lower median timescale ($0.38 \, {\rm Gyr}$ in TNG100 compared to $0.47 \, {\rm Gyr}$ in \astridN). Interestingly, in both TNG100 and \astridN, the quenching timescale has no significant correlation with the stellar mass.

In Fig.~\ref{fig:Astrid-QG-example}, we present an illustration of a massive galaxy quenching in \astridN. This example galaxy quenches at $z=3$, like the TNG100 example galaxy, and it has a similar stellar mass at this redshift. The mass of its central SMBH at quenching is $\logMBH = 8.7$. As in Fig.~\ref{fig:TNG100-QG-example}, we map the galaxy's gas temperature and stellar age. We display snapshots from $z = 4$ to $z = 3$ (a period of $0.61 \, {\rm Gyr}$). Quenching occurs at the final snapshot. At the galaxy's center, we again observe that the gas is heated, but there are no significant gas cavities.

\subsection{Environment of quenched galaxies} \label{sec:environment}

\begin{figure}
    \includegraphics[width=\columnwidth]{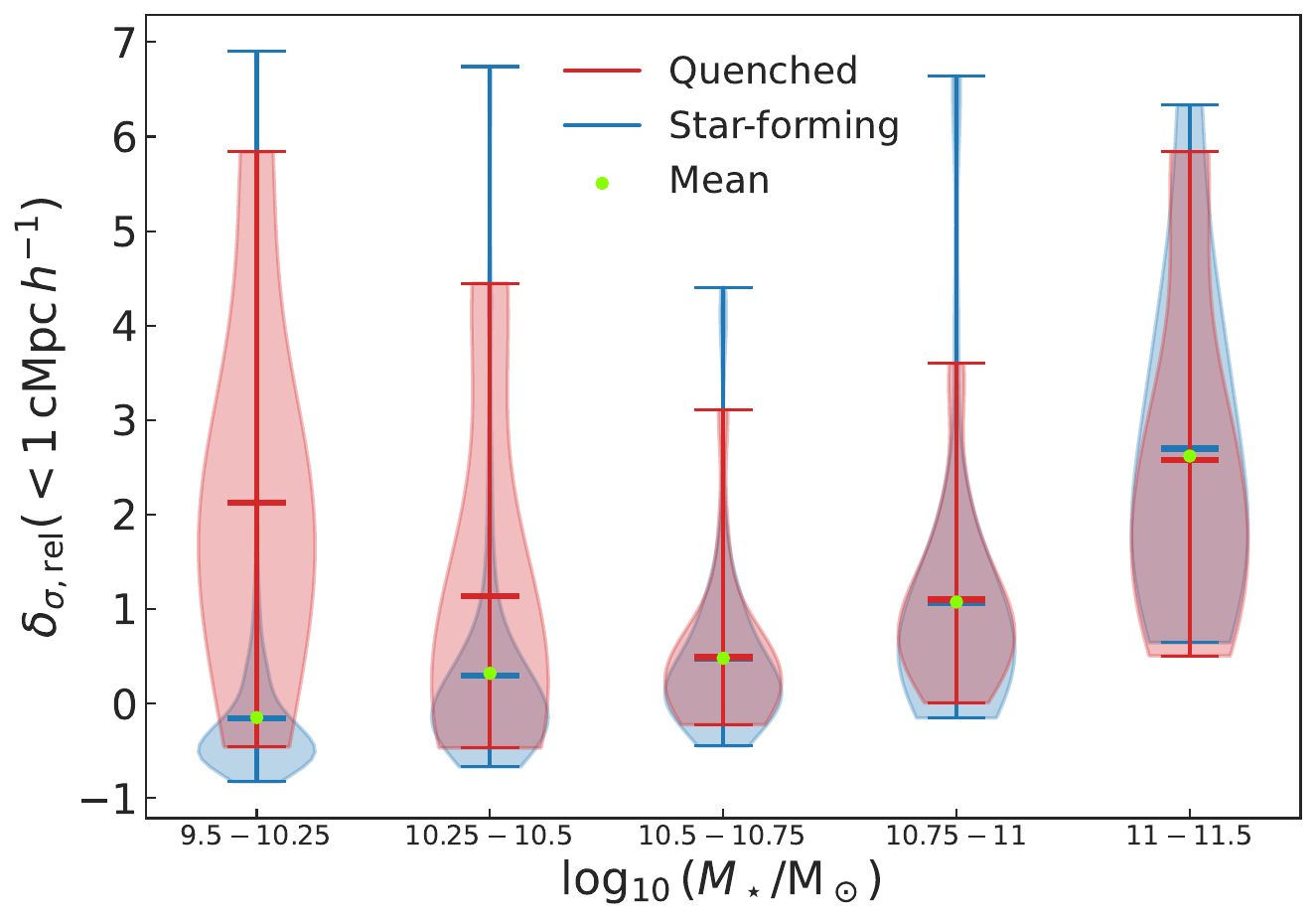}
    \caption{Distribution of the deviation from the mean density contrast, $\delta_{\sigma,\rm{rel}}(<~1~\distMpc)$, for quenched and star-forming massive galaxies at $z=3$ divided into stellar mass bins. The horizontal lines mark the minimum, mean, and maximum values of the distributions. At lower stellar masses, the quenched galaxies tend to lie in higher local overdensities than their star-forming counterparts. For each stellar mass bin, the mean value of $\delta_{\sigma,\rm{rel}}(<~1~\distMpc)$ among both the quenched and star-forming galaxies is marked in red. We see that this mean value increases with stellar mass.}
    \label{fig:density_violinplot}
\end{figure}

To quantify the environment of high-redshift quenched massive galaxies compared to their star-forming counterparts, we use equations from \cite{Kimmig_2023}. Note that we focus here on galaxies in TNG100. The density contrast is given by:
\begin{equation}
    \delta(<R) = \frac{\rho(<R) - \bar{\rho}}{\bar{\rho}},
\end{equation}
where $\rho(<R)$ is the mass density within a sphere of radius $R$ centered on the galaxy (calculated by summing the total mass of all galaxies whose centers lie in the sphere and dividing by the comoving volume of the sphere), and $\bar{\rho}$ is the mean density at that redshift (calculated by summing the total mass of all galaxies in the simulation and dividing by the comoving simulation volume). Then, the deviation from the mean density contrast is given by:
\begin{equation}
    \delta_{\sigma,\rm{rel}}(<R) = \frac{\delta(<R) - \bar{\delta}(<R)}{\sigma_{\delta,R}},
\end{equation}
where $\bar{\delta}(<R)$ is the mean value of $\delta(<R)$ for all of the galaxies in our sample (both quenched and star-forming), and $\sigma_{\delta,R}$ is the standard deviation.

In Fig. \ref{fig:density_violinplot}, we show the distribution of $\delta_{\sigma,\rm{rel}}(<~1~\distMpc)$ for quenched and star-forming galaxies at $z=3$, divided into stellar mass bins ranging from $\logMstar = 9.5 - 11.5$. While not shown here, we found similar distributions at $z=2$. In the lower stellar mass bins, we see that the quenched galaxies tend to lie in significantly denser local environments than the star-forming galaxies. This supports our claim that the environment plays a role in quenching the lower-mass high-redshift galaxies. As noted in Sec.~\ref{sec:intro}, environmental quenching occurs when galaxies are located in large clusters where the hot galactic halo is impenetrable to cold gas streams (e.g. \citealt{Gunn_1972, Dressler_1980, Mihos_1996, Kawata_2008, vdB_2008, Peng_2010, Donnari_2021_mechanisms}).

We also see in Fig. \ref{fig:density_violinplot} that the mean value of $\delta_{\sigma,\rm{rel}}(<1\distMpc)$ increases with stellar mass. We note that, as a result, the quenched galaxies have a higher average local overdensity than the star-forming galaxies (when considering the entire studied mass range), because their stellar masses are larger on average.

This is in agreement with \cite{Kurinchi_2023}, who also study quenched massive galaxies at $z \gtrsim 3$ in \tngN, and find that they tend to lie in more massive halos and denser regions. Interestingly, these results contradict \cite{Kimmig_2023}, who find that high-$z$ quenched galaxies in the \texttt{Magneticum Pathfinder} simulation tend to lie in local underdensities compared to star-forming galaxies with comparable masses ($\logMstar > 10.5$). This may be due to the difference in AGN feedback prescriptions (see \citealt{Steinborn_2015} for the prescription used in \texttt{Magneticum Pathfinder}). \cite{Szpila_2024} find that high-$z$ quenched galaxies in the \texttt{SIMBA-C} simulation typically form in overdense environments, but by the time of quenching, their environments are similar to those of star-forming galaxies.

\subsection{Post-quenching evolution} \label{sec:BCGs}

\begin{figure}
    \includegraphics[width=\columnwidth]{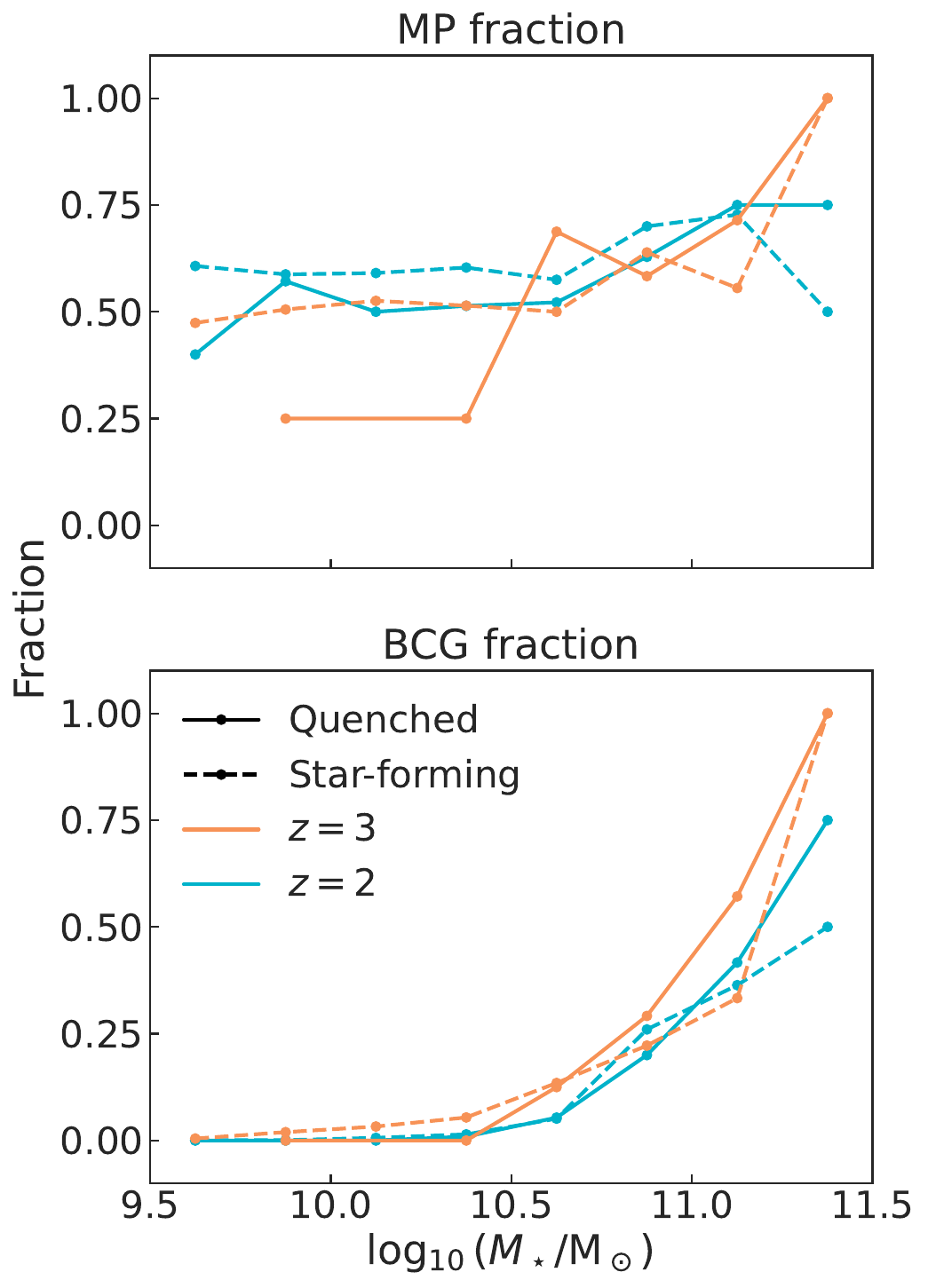}
    \caption{MP and BCG fractions vs. stellar mass for quenched and star-forming massive galaxies in TNG100 at $z=3$ and $z=2$. The MP and BCG fractions refer to the fraction of galaxies that are the main progenitor of a $z=0$ galaxy or, specifically, a $z=0$ BCG, respectively. Within a given stellar mass bin, the MP and BCG fractions are similar for the quenched and star-forming galaxies. More massive galaxies are significantly more likely to be the main progenitor of a $z=0$ BCG.}
    \label{fig:mp_frac}
\end{figure}

To investigate how galaxies evolve after quenching, we analyze the fraction of quenched and star-forming massive galaxies in TNG100 that serve as the main progenitors of their $z=0$ descendants and the fraction of galaxies that are the main progenitors of $z=0$ brightest cluster galaxies (BCGs). We refer to the first quantity as the MP fraction and the second as the BCG fraction. We define a galaxy to be a BCG if it meets the following criteria: (i) it is the ``primary'' subhalo in its halo (largest number of bound particles/cells); (ii) it has the largest stellar mass in the halo; (iii) the halo has more than 100 subhalos with nonzero stellar mass. The third criterion selects the most massive halos (and by extension, the most massive primary subhalos). For example, we calculate at $z=0$ the mean and standard deviation of the logarithm of the halo mass, $\log_{10}{(M_{halo}/{\rm M}_\odot)}$. For the BCG host halos identified by our criteria, the mean and standard deviation are $13.71$ and $0.35$, respectively, compared to $8.55$ and $0.53$ for all the halos in the simulation.

As in Fig.~\ref{fig:feedback_frac}, we compute the MP and BCG fractions at $z=3$ and $z=2$ within stellar mass bins of width $0.25 \, \rm dex$. Our results are presented in Fig.~\ref{fig:mp_frac}. The MP fraction shows no clear trend between quenched and star-forming galaxies and between redshifts, but there does appear to be some stellar mass dependence at $z=3$. The BCG fraction is similar for both quenched and star-forming galaxies but is higher at $z=3$ than at $z=2$ for the largest stellar masses. Most noticeably, the BCG fraction strongly depends on stellar mass: more massive galaxies are significantly more likely to be the main progenitor of a $z=0$ BCG. We note that our MP fractions in Fig.~\ref{fig:mp_frac} are in contrast with the finding in \cite{Hartley_2023} that only 1 of the 5 earliest quenched massive galaxies in TNG300 ($10.5 < \logMstar < 11$) is the main progenitor of its $z=0$ descendant. However, we note the following differences between our work and \cite{Hartley_2023}: (i) they use TNG300, while we use TNG100; (ii) they use galaxies that quench by $z=4.2$, while we study galaxies that quench at lower redshifts, i.e., by $z=3$ or $z=2$; (iii) only 5 galaxies meet their criteria, so their lower MP fraction may simply be a result of the small sample size.

While we do not see any significant difference in the BCG fraction between quenched and star-forming galaxies when accounting for stellar mass differences, we note that because the quenched galaxies have larger average stellar masses within the studied mass range, they have a higher overall BCG fraction. At $z=3$, the MP and BCG fractions for quenched galaxies are $60\%$ and $30\%$  , respectively, compared to $50\%$ and $3\%$ for star-forming galaxies. At $z=2$, these values are $56\%$ and $11\%$ for quenched galaxies and $60\%$ and $1\%$ for star-forming galaxies.

Many of the $z=0$ galaxy main progenitors that are quenched at high redshift are later reactivated: their sSFR rises back above (or near) $10^{-10} \, {\rm yr}^{-1}$. Returning to Fig.~\ref{fig:evolution_plot_fdbk}, we see that the TNG100 example galaxy reactivates at $z \sim 1$; this galaxy is the main progenitor of a $z=0$ BCG. 
In the last column of Fig.~\ref{fig:TNG100-QG-example}, we observe surrounding galaxies merging onto the central example galaxy at $z=1$, bringing in new gas and reactivating star formation. 

To quantify this reactivation, we take the quenched massive galaxies at $z=3$ in TNG100 and use merger trees to track their sSFR forward in time. If a galaxy is the main progenitor of a $z=0$ galaxy, we track its evolution all the way to its $z=0$ descendant. Otherwise, we stop when it merges with a more massive galaxy. We find that $64\%$ of the galaxies that are the main progenitor of a $z=0$ BCG have $\logsSFR > -10$ at some point after $z=3$, compared to $47\%$ of galaxies that are the main progenitor of any $z=0$ galaxy and $21\%$ of galaxies that are not. The overall fraction among all the galaxies is $38\%$. The reactivation does not last: only $4\%$ of the $z=0$ descendants of the galaxies have $\logsSFR > -10$. Similar behavior has been seen in the \texttt{SIMBA-C} simulation: $\gtrsim 30 \%$ of high-redshift quenched massive galaxies are reactivated, but this fraction drops quickly at $z \lesssim 2$ \citep{Szpila_2024}.

\section{Summary and Conclusions} \label{sec:discussion}
This study is motivated by recent JWST observations providing demographic estimates for the population of quenched massive galaxies at $3 < z < 5$ (see, e.g., \citealt{Carnall_2023,Long_2023,Valentino_2023, Gould_2023}).
We used the cosmological simulations \tng and \astrid to investigate the rise of these galaxies. Our primary findings are summarized below:

\begin{itemize}
    \item \textbf{Abundance of quenched galaxies in simulations:} \tng and \astrid significantly underestimate the comoving number density of quenched massive galaxies at $z \gtrsim 3$ compared to JWST data; this discrepancy reaches $>2$ orders of magnitude at the high end of the redshift range that we investigated. 
    \item \textbf{Role of AGN feedback:} AGN feedback is identified as the key quenching mechanism for massive galaxies in both simulations. In \tngN, the kinetic mode of AGN feedback is predominant, rapidly quenching massive galaxies. In contrast, \astrid relies on AGN thermal feedback at $z>2.3$, which, although less effective, is still key to quenching on a slightly longer timescale. We highlight, however, that the simulation suites used differ in many aspects, making an absolute comparison of the role of kinetic feedback challenging.
    \item \textbf{Quenching timescales:} Massive galaxies at $z>3$ have typical quenching timescales of $\sim 200-600$ Myr.
    \item \textbf{Characteristics of quenched galaxies:} In both \tng and \astridN, black holes in quenched massive galaxies are overmassive compared to the average $M_{\rm BH} - M_{\star}$ scaling relation. This suggests that AGN feedback is more pronounced in these galaxies, contributing to their quiescence. Additionally, galaxies that quench earlier tend to be more massive and accumulate stellar mass more rapidly.
    \item \textbf{Environments of quenched galaxies:} Lower-mass high-redshift quenched galaxies tend to have significantly denser local environments than their star-forming counterparts, suggesting that the environment plays an important role in quenching lower-mass galaxies. The average local overdensity also increases with stellar mass.
    \item \textbf{Post-quenching evolution:} A notable fraction of quenched massive galaxies experience reactivation of star formation at later stages, primarily due to interactions and mergers that bring fresh gas into the galaxy. These galaxies often evolve to become the primary progenitors of $z=0$ BCGs, indicating a significant evolutionary pathway from high-redshift quiescence to present-day galaxy clusters.
\end{itemize}

Our results emphasize the need for enhanced physical models of AGN feedback to better match observational data, particularly at high redshifts. Future observations from JWST, combined with improved simulations, are expected to refine our understanding of the formation and evolution of massive galaxies in the early Universe.

\vspace{10pt}
\noindent \textit{Acknowledgments:}
We thank the referee for providing insightful comments on the paper.
We thank Erica Nelson and Arianna Long for helpful discussions.
E.W. acknowledges that this material is based upon work supported by the National Science Foundation Graduate Research Fellowship under Grant No. DGE-2139841 and by the Harvard College Research Program (HCRP). This work was completed in part as a class project for Astronomy 91R, taught at Harvard College by Charles Alcock. 
F.P. acknowledges support from a Clay Fellowship administered by the Smithsonian Astrophysical Observatory. 
Y.N. acknowledges support from an ITC postdoctoral fellowship by Harvard University. 
L.H. acknowledges support by the Simons Collaboration on ``Learning the Universe.''



\bibliography{ms}{}
\bibliographystyle{aasjournal}


\end{document}